\documentclass[11pt]{article}

\usepackage{cite}
\usepackage{amssymb}
\usepackage{amsmath}
\usepackage{bm}
\usepackage{graphicx}
\usepackage{epstopdf}
 \usepackage [latin1]{inputenc}

\begin{document}

\title{Stapp, Bohm and the Algebra of Process.}
\author{B. J. Hiley\footnote{E-mail address b.hiley@bbk.ac.uk.}.}
\date{TPRU, Birkbeck, University of London, Malet Street, \\London WC1E 7HX.\\Physics Department, University College London, Gower Street, London WC1E 6BT. }
\maketitle

\begin{abstract}

Henry Stapp has made many significant contributions in quantum physics and its use in trying to understand the mind-matter relationship.  I have been influenced by his use of the notion of {\em process} to bring more clarity to understand 
quantum phenomena.  In this paper I want to summarise the latest ideas on the time development of quantum processes that relate the transformation theory of Dirac, Feynman and Schwinger to the notion of weak values  which has triggered  experimental investigations of the nature of a deeper underlying stochastic structure of quantum processes.

\end{abstract}

\section{Introduction}

It is a great privilege  to contribute to Henry Stapp's ninetieth birthday Festschrift.  My interactions with Henry go back to the seventies when I was starting out on my exploration of quantum mechanics.  I joined David Bohm at Birkbeck College, London in 1961, beginning a collaboration that lasted for thirty-five years.  In those early years, we did not discuss his famous 1952 papers~\cite{db52} on `hidden variables' which eventually morphed into `Bohmian mechanics', the latter was an approach that we had both rejected from its inception, instead outlining an alternative which we summarised in ``The Undivided Universe"~\cite{dbbh93}.  We were exploring a more radical idea that Bohm called `structure process'~\cite{db65}.  The aim was to provide a fundamentally new approach to quantum phenomena which we hoped would ultimately lead to a theory of quantum gravity.

We were joined by Roger Penrose who at the time was in the Birkbeck mathematics department, developing a theory of spin networks and twistor theory~\cite{rp67, rp71}.   Naturally our discussions centred around 
the question of how to unite QM and GR.  In this environment my introduction to quantum theory was unorthodox to say the least! 

The approaches of both Bohm and Penrose abandoned the idea of assuming an {\em a priori} given space-time in which fields and particles evolve in time. Instead we were exploring possible deeper structures from which space-time itself would emerge as some form of coarse grained approximation.  Penrose showed how the  classical rotation group emerged from a  quantum spin structure when large quantum numbers were involved.  It is this structure that now forms the basis of quantum loop gravity, a subject that has recently developed rapidly into the type of theory that we had in mind way back in the sixties but had failed to make any headway at that stage.

But I get ahead of myself.  In the sixties we first discussed a possibility of describing  `structure process' in terms of an underlying simplicial structure using a discrete de Rham cohomology to provide a link to quantum numbers~\cite{dbbh70}.  However the model was too static 
emphasising more the `structure' at the expense of the `process'.  I then discovered the work of Benn and Tucker~\cite{bt87}, who showed how 
differential forms used in the de Rham approach could be generalised to link with the abstract generators of Clifford algebras.  The vital piece of background to this algebra was the work of  Clifford himself.  He had started from the idea of process, which depended crucially on the order of action. Remarkably Clifford~\cite{wc82} was working, in those pre-quantum days, entirely within classical physics, yet discovered an algebra that now plays a key role in quantum mechanics when spin and relativity are introduced.  In light of this, we introduced the notion of the  `algebra of process'.   This provided a key link with Penrose's twistors which were, of course, the semi-spinors of the conformal Clifford algebra (see Bohm and Hiley~\cite{dbbh84}).

\section{The Emergence of the Classical World}

\subsection{Gentle Photons}

It was at this time that we became aware of the ideas that Henry and his colleague, Geoff Chew,  were exploring at Berkeley.  There was a common theme.  We all agreed that quantum theory in the hands of Bohr offered a set of rules for calculating the statistical results under well-defined experimental conditions.  However it did not provide an ontology that would unambiguously remove the observer from playing an essential role in the process.  Henry proposed that it should be the electromagnetic field that provided the link between classical and quantum properties, a link in which the twistor was to play a role.  My preference is for  gravity to play such a role since it permeates everything but that is for the future.

Following the ideas discussed in an early paper of Bloch and Norsdsieck~\cite{bn37}, 
Henry suggested that the `infrared catastrophe', rather than being a problem, should be used in a positive way to provide a method to separate out the classical aspect of the total process.  In this way he~\cite{hs83} 
 completely solved the technical problem and showed that it was the coherent state of the electromagnetic field that replaced the `observer' in  Bohr's approach.  Recall that coherent states have classical-like solutions in phase space.  
In the case of the em field, the expectation value in the coherent state $|A(t)\rangle$ of the quantum operator $\hat A(x)$ corresponding to the vector potential can be written as 
\begin{eqnarray*}
A(x)=\langle A(x')|\hat A(x)|A(x')\rangle.
\end{eqnarray*}
Further work shows that the $S$-matrix can be expressed in terms of $A(x)$ establishing the presence of the classical electromagnetic field in the theory with its position in space-time, rather than configuration space.  Furthermore the incorporation of light into the $S$-matrix automatically brings into this description an exact classical level  that is coordinated to the ordinary four-dimensional space-time continuum of special relativity.

By this means the structure process can provide a well-ordered sequence of actual events in space-time so that it is meaningful to regard each quantum process as a sequence of actual events in space-time.  Thus our ontology contains no explicit dependence on human observers.  In this sense the ideas discussed in this paper are different from the position Henry now favours.   For example, the Feynman path can be considered as an actual sequence of events in space-time, totally independent of human intervention.  This enables us to show exactly how the Bohm approach that I had worked on in the seventies fits into the standard approach to quantum mechanics~\cite{rfbh18}.

\subsection{The Bohm Approach}

In the seventies I was encouraged by two of our research students, Chris Philippidis and Chris Dewdney, to examine in more detail Bohm's 1952 papers~\cite{db52}.  The titles of these papers contained the phrase `hidden variables', an approach that many thought had failed in its aims, including myself and I had never taken it seriously.  In spite of this, we decided to use the rapidly developing new computer technology to calculate `trajectories', examine their form and look at the detailed structure of the quantum potential [see equation (\ref{eq:QHJ}) below] for various characteristic quantum phenomena such as two-slit interference, barrier penetration, scattering by square wells etc.  

The approach was simple, take the classical canonical relations $p=\nabla S$ and $E=-\partial_t S$, replace the classical action by the phase of the wave function and calculate `trajectories'.    Amazingly we found that the results gave a powerful intuitive picture of what {\em could} be going on provided our assumptions were correct.  The key question was why was this approach apparently working so well?

It was much later when I noticed that Dirac~\cite{pd47} had introduced an algebraic approach that turns out to
 be the forerunner of the Bohm approach~\cite{db52}, producing exactly the same equations that Bohm used.  However Dirac argued that proceeding in this way the existence of a local momentum would violate the uncertainty principle.  Bohm, in contrast, showed that this was not true.  Furthermore he realised that one could still retain classical ideas by using the first order WKB approximation, but keeping all the terms of the expansion demanded a radical change of outlook.  It was those terms that summed into the quantum potential, a notion that Heisenberg regarded as {\em ad hoc}~\cite{wh58}.  However it was the forerunner of deformation quantum mechanics~\cite{fbmf78}

 Dirac went on to suggest that the Lagrangian played a key role in his approach and so I took the Lagrangian that Heisenberg had used to `derive' the Schr\"{o}dinger equation.  Rewriting the Lagrangian using the polar decomposition of the wave function $\psi=Re^{iS/\hbar}$ one finds that the Euler-Lagrange equations give the two equations that Bohm had used, namely,
\begin{eqnarray}
\frac{\partial P}{\partial t}+\nabla.\left(P\frac{\nabla S}{m}\right)=0 \label{eq:ConP}
\end{eqnarray}
and
 \begin{eqnarray}
\frac{\partial S}{\partial t} +\frac{(\nabla S)^2}{2m} +Q +V=0 \label{eq:QHJ}
\end{eqnarray}
where $Q=-\frac{\hbar^2}{2mR}(\nabla^2 R)$ is the expression for the quantum potential.  But this still does not explain why the relation $p=\nabla S$  works. 

 Once we have the Lagrangian, we can use the general expression for the energy momentum tensor to find
\begin{eqnarray*}
T^{0\mu}=\frac{i}{2}[\psi^*{\overleftrightarrow\partial^\mu}\psi]=-\rho\partial^\mu S.
\end{eqnarray*}
Explicitly $p=\nabla S=T^{0j}/\rho$ and $E=-\partial_tS=T^{00}/\rho$.  So the Bohm momentum actually emerges from the energy-momentum tensor derived from the Schr\"{o}dinger Lagrangian. But there is more.  The trace of the energy-momentum is
\begin{eqnarray*}
(T^{kk}-{\cal L}\delta^{kk})/\rho=\frac{(\partial^kS)^2}{2m}+\frac{(\partial^kR)^2}{2mR^2} +V.
\end{eqnarray*}
Thus not only does the kinetic energy, $KE_B=p^2/2m=(\nabla S)^2/2m$, emerge, but there also appears a new form of kinetic energy, namely, $KE_O = (\nabla R)^2/2mR^2$.   This latter is clearly connected with the appearance of the quantum potential.  It turns out that I had rediscovered some early work of Takabayasi~\cite{tt54}.

I found all this very reminiscent of early work of Feynman~\cite{rpf48} and Schwinger~\cite{js51, js53a, js53} when they laid the foundations of quantum field theory.
 Schwinger~\cite{js53} argued that the fundamental quantum dynamical laws would find their proper expression in terms of transition  amplitudes [TAs], not in terms of the Schr\"{o}dinger wave functions.  
The key ingredient in his work was the energy-momentum tensor which he used to define momentum TA.  In the non-relativistic theory, the momentum TA is simply
\begin{eqnarray}
\langle \hat P^\mu\rangle=\frac{\langle \phi(x,t)|\hat P^\mu| \psi(x_0,t_0)\rangle}{ \langle \phi(x,t)|\psi(x_0,t_0)\rangle}	\label{eq:TPA}
\end{eqnarray}
an expression that is exactly the same as the weak value of the momentum, about which we will have more to say later.

It was Dirac~\cite{pd45} who suggested that, in the non-relativistic case, we should divide the connection between two states $\psi(x_0, t_0)$ and $\phi(x, t)$ into a series of infinitesimal time steps $\epsilon=t_{j+1}-t_j$, enabling us to construct a `path'  out of a series of  TAs so that
\begin{eqnarray}
\langle x_t|x_{t_0}\rangle=\int \langle x_t|x_j\rangle dx_j\langle x_j|x_{j-1}\rangle\dots\langle x_2|x_1\rangle dx_1\langle x_1|x_{t_0}\rangle
\label{eq:DTraj}
\end{eqnarray}
where the $\langle x_{j+1}|x_j\rangle$ are a set of infinitesimal TAs.  In this way, as Dirac argues~\cite{pd45}, we can discuss trajectories for the
 motion of a quantum particle, which makes quantum mechanics more closely resemble classical mechanics. Indeed the method enables one to bring out the close analogy between classical and quantum contact transformations, an analogy that Bohm highlights in his book, ``Quantum Theory"~\cite{db51}.

It is interesting to note that in sections 31 and 32 of his book, Dirac~\cite{pd47} derived equations (\ref{eq:ConP}) and (\ref{eq:QHJ}) from an algebraic point of view but did not pursue the approach because he thought the uncertainty principle would be violated.  Bohm showed that this conclusion was not correct by developing his causal approach\footnote{ Bohm refers specifically to these two sections in his book~\cite{db51} and  so he was well aware of what Dirac had done.}.  All this prompts the question `Is there a relation between the Feynman paths and the Bohm trajectories?'

\section{Feynman Paths and Bohm Trajectories}

In equation (\ref{eq:DTraj}) one writes $\langle x|x'\rangle=\exp[iS(x,x')]$, where Feynman assumes $S(x,x')=\delta\int{\cal L}(x, x')dt$, the classical action and therefore 
\begin{eqnarray*}
S(x,x')=\frac{m(x-x')^2}{2\epsilon}
\end{eqnarray*}
$\epsilon$ being a small time interval.  The momentum TPA at a point $X$ between $x'$ and $x$ is
\begin{eqnarray*}
p_X(x,x')=\frac{\partial S(X, x')}{\partial X}+\frac{\partial S(x, X)}{\partial X}=\left[\frac{(X-x')}{\epsilon}-\frac{(x-X)}{\epsilon}\right].
\end{eqnarray*}
Notice that the derivative is not continuous at $X$.  Instead we have a `backward derivative' $(X-x')/\epsilon$ and a `forward derivative'  $(x-X)/\epsilon$ at  $X$.  Thus the Feynman path is continuous but nowhere differentiable.

 Over time an ensemble of individual particles pass through $X$, so that there is a distribution of momenta arriving at $X$ and a distribution of momenta leaving the point. Thus at each point $X$, we have an average value of the momentum and that average value must be determined by the wave function.  The average momentum at a point turns out to be the Bohm momentum $p=\nabla S$, $S$ being the phase of the wave function.
 
 \begin{figure}[htbp] 
    \centering
    \includegraphics[width=3in]{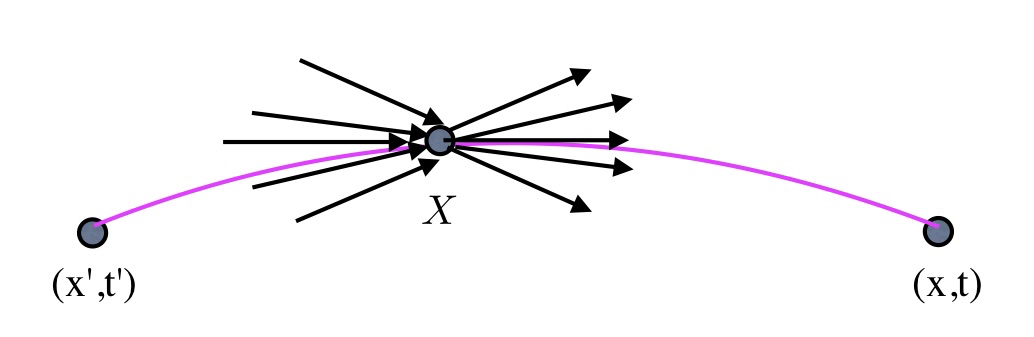} 
    \caption{Enfolding and unfolding at a point}
    \label{fig:spray}
 \end{figure}

To see how this result follows, consider the process shown in Figure \ref{fig:spray}. This gives an image of an ensemble of actual individual quantum events occurring at $X$ together with the incoming and outgoing  sprays of momenta. Thus we have two contributions to consider, one spray coming from the point $x'$ and the other spray leaving for the point $x$.  We must determine the distribution of momenta in each spray to produce a result that is consistent with the wave function $\psi(X)$ at $X$.     Therefore we can write
\begin{eqnarray*}
\lim_{x'\rightarrow X}\psi(x')=\int\phi(p')e^{ip'X} dp'\hspace{0.5cm}\mbox{and}\hspace{0.5cm} \lim_{X\rightarrow x}\psi^*(x)=\int \phi^*(p)e^{-ipX}dp.
\end{eqnarray*}
The $\phi(p')$ contains information regarding the probability distribution of the incoming momentum spray, while $\phi^*(p)$ contains information about the  probability distribution in the outgoing momentum spray.  These wave functions must be such that in the limit $\epsilon\rightarrow 0$ they are consistent with the wave function $\psi(X)$.

Thus we can define the mean momentum, $\overline{\overline {P}}(X)$ as
\begin{eqnarray}
\rho(X)\overline{\overline {P}}(X)=\int\int P\phi^*(p)e^{-ipX}\phi(p')e^{ip'X}\delta(P-(p'+p)/2)dPdpdp'		\label{eq:MMPP}
\end{eqnarray}
where $\rho(X)$ is the probability density at $X$.   We have added the restriction $\delta(P-(p'+p)/2)$ since momentum is conserved at $X$.  We can rewrite equation (\ref{eq:MMPP}) and form
\begin{eqnarray*}
\rho(X)\overline{\overline {P}}(X)=\frac{1}{2\pi}\int\int P\phi^*(p+\theta/2)e^{-iX\theta}\phi(p-\theta/2)d\theta dP
\end{eqnarray*}
or equivalently taking Fourier transforms
\begin{eqnarray*}
\rho(X)\overline{\overline {P}}(X)=\frac{1}{2\pi}\int\int P\psi^*(X-\sigma/2)e^{-iP\sigma}\psi(X+\sigma/2)d\sigma dP
\end{eqnarray*}
which means that $\overline{\overline {P}}(X)$  is the conditional expectation value of the momentum weighted by the Wigner function.
Equation (\ref{eq:MMPP}) can be put in the form
 \begin{eqnarray}
    \rho(X)\overline{\overline {P}}(X)=\left(\frac{1}{2i}\right)[(\partial_{x_1}-\partial_{x_2})\psi(x_1)\psi(x_2)]_{x_1=x_2=X}	\label{eq:MMXX}
  \end{eqnarray}
  an equation that appears in the Moyal approach~\cite{jm49}, which is
   based on a different but isomorphic non-commutative algebra. 
If we evaluate this expression for the wave function written in polar form $\psi(x)=R(x)\exp[iS(x)]$, we find  $\overline{\overline P}(X)=\nabla S(X)$ which is just the Bohm momentum. 

Since the Bohm momentum is an average value, the trajectories calculated from them must be averages, so that each Bohm  `trajectory' is an average of an ensemble of individual Feynman paths.    It is not the momentum of a single `particle' passing the point $X$, as assumed in Bohmian mechanics, but the mean {\em momentum flow} at the point in question.

\section{Weak Values}

The previous section was motivated by the appearance in 2011  of an experiment reporting the construction of `photon trajectories' using a two-slit optical setup that  measured the weak values of the transverse momentum~\cite{skbbsr11}. This was all made  possible by utilising a general idea introduced by
 Aharonov, Albert and Vaidman~\cite{yalv88} who defined the weak value of any
  operator $\hat A$ by 
\begin{eqnarray*}
\langle \hat A\rangle=\frac{\langle \phi(t)|\hat A |\psi(t_0)\rangle}{ \langle \phi|(t)|\psi(t_0)\rangle}.
\end{eqnarray*}
Clearly equation (\ref{eq:TPA}) is a special case of a weak value.  Indeed
 Leavens~\cite{rl05} and Wisemen~\cite{hw07} argued that the weak 
 value of the momentum provided, in principle, a way of experimentally constructing Bohm trajectories.  It was this connection that Kocsis {\em et al.}~\cite{skbbsr11} exploited in their two-slit optical experiment.  
 Their results were remarkably similar to the Bohm trajectories~\cite{cpcd79}. (For a detailed comparison see  Coffey and Wyatt~\cite{tcrw11}). 
These experimental results show clearly the statistical nature of the results used to construct the flow lines, thus confirming the statistical nature of the Bohm trajectories.

There is a problem for the claimed interpretation of flow lines  in that photons, per se,  do not have trajectories.   Nevertheless Flack and Hiley~\cite{rfbh16} showed that what Kocsis {\em et al.}
~\cite{skbbsr11} had constructed were mean momentum flow lines using  the real part of the {\em weak} Poynting vector.  If one requires a more appropriate comparison to the Bohm trajectories then one should experiment using atoms.  In fact our group at UCL are at present measuring weak values of  momentum using argon and helium~\cite{jmpe16,vmrf17} and the experiments are very close to completion. 

\section{The Algebra of Process}
\subsection{The Technical Details}

It should by now be apparent that the Bohm approach has its origins in the non-commutative algebra pioneered by Born, Heisenberg and Jordon~\cite{bhj} and which I have called `the algebra of process'~\cite{bh80}.  The key link  appears in
 Dirac's  ``The Principles of Quantum Mechanics"~\cite{pd47} .  There a symbol, $\;\rangle$, the Ôstandard ketÕ is introduced into the algebra. This enables the wave function $\psi$ to be written as a {\em wave operator}, $\psi( \hat Q,\hat P)\;\rangle$; technically the new object turns the wave function into an element of a left ideal.  To complete the algebra, a dual symbol, the standard bra is introduced.  Thus Dirac has replaced the usual Hilbert space by a non-commutative algebra formed by the symbols $[\hat Q,\hat P, \epsilon]$ where the $\epsilon=\;\rangle\langle\;$, an idempotent\footnote{The idempotent symbol is introduced because the Heisenberg algebra is nilpotent and contains no idempotent.}.  In this way all the essential details of the quantum process are contained in the algebra.  

In effect what Dirac has done is to introduce a new representation, which I call the Dirac-Bohm picture~\cite{bhgd18}.  This representation is unitarily equivalent to the Schr\"{o}dinger picture and supplements the Heisenberg, Interaction and Fock pictures. It is different in that the unitary transformation is based on the action rather than the energy.  Indeed $\epsilon$ plays an analogous role to the vacuum projector in the Fock representation.

Using this approach I was able to propose two time evolution equations within the algebra~\cite{bh15}, equations implicitly contained in Dirac's 
work~\cite{od92}.  A similar pair of equations can also be constructed
 in the Moyal algebra.  I have used these equations to extend the Dirac-Bohm picture to Clifford algebras and shown how the Pauli and Dirac equations fit into the structure~\cite{bhbc12}.  Contrary to the usual 
 assumption, there is no problem with extending this picture to include relativity. In passing I should also like to mention that this approach was inspired by Penrose's development of twistors.  It is this algebra that provides a possible approach to quantum gravity.
 
 \subsection{The Overarching Philosophy}
 
 The theme of this paper was motivated by the paper  that Henry kindly submitted to Bohm's 70th Festschrift~\cite{bhdp87}.  There Henry talks
about the `Bohm-Heisenberg idea of events'.  I agree on the notion of quantum events that actualise, but I wanted to clarify how Bohm's `52 paper~\cite{db52} related to the Heisenberg non-commutative algebra.  The Bohm approach emerges as a  coarse grained average, giving the appearance of a deterministic approach, but being, in fact, very different from classical determinism.  

Our approach restores the position-momentum symmetry and so two views emerge, a phase space constructed from $(x, p=\nabla_xS(x))$ or from $(x=-\nabla_p S(p), p)$.  These are the shadow phase spaces.  In the Dirac picture they correspond to choosing the idempotent defined by $\hat P\epsilon_x=0$ or $\hat X\epsilon_p=0$\footnote{ In Fock space these are analogous to $a|0\rangle=0$ or $a^\dag|F\rangle=0$ where $|F\rangle$ is the full or plenum state. The latter is more commonly experienced with fermions.}.

Bohr proposed that we understand this dual view through the Principle of Complementarity, a philosophical principle that does not sit comfortably with physicists in general.   Bohm proposed a new notion of the implicate-explicate order.  The need for non-commutativity suggests that we can no longer provide one unique, God's-eye, view of natural phenomena.  Because we are inside, as it were, we can only project out partial views determined by the experimental conditions which enable us to construct  particular shadow manifolds, or explicate orders.  The underlying reality is implicate.  Bohm  investigated the consequences of this implicate view of reality, not only in physics but in other areas of intellectual discourse.  The one that would most interest here is the application of these ideas to mind, but I don't have the space to discuss this here.  Have a happy 90th Henry!



\bibliography{myfile}{}

\begin{thebibliography}{99}

\bibitem{db52} Bohm, D., A Suggested Interpretation of the Quantum Theory in Terms of Hidden Variables, I, {\em Phys. Rev}., {\bf 85} (1952) 166-179; and II, {\bf 85}, (1952), 180-193.

\bibitem{dbbh93} Bohm, D. and Hiley, B. J., The Undivided Universe: an Ontological Interpretation of Quantum Theory, Routledge, London 1993.

\bibitem{db65} Bohm, D., Problems in the Basic Concepts of Physics, in {\em Satyendranath Bose 70th Birthday Commemoration Volume} Part II, pp. 279-318, 1965.

\bibitem{rp71}Penrose, R., Angular Momentum: a Combinatorial approach to Space-time, in {\em Quantum Theory and Beyond}, ed. Bastin, T., Cambridge University Press, Cambridge, 151-180, 1971.

\bibitem{rp67}Penrose, R., Twistor Algebra, {\em J. Maths Phys}., {\bf 8}, (1967), 345-366.

\bibitem{dbbh70} Bohm, D., Hiley, B. J. and  Stuart, A.E.G.,  On a New Mode of Description in Physics,   Int. J. Theor. Phys. 3 (1970) 171-183.

\bibitem{bt87} Benn, I. M. and Tucker, R. W., {\em An Introduction to Spinors and Geometry with Applications in Physics}, Adam Hilger, 1987.

\bibitem{wc82}Clifford,  W.K., {\em Mathematical Papers}, XLII, Further note on biquaternions, 385-94, Macmillan, London, 1882.

\bibitem{dbbh84} Bohm, D. J. and Hiley, B. J.,  Generalization of the Twistor to Clifford Algebras as a Basis for Geometry,    Revista Braseilra de Fisica, Vol. Especial Os 70 anos de Mario Sch\"{o}nberg, 1-26, (1984).

\bibitem{bn37} Bloch, F. and Nordsieck, A., Notes on the Radiation Field of the Electron, {\em Phys. Rev.,} {\bf 52} (1937) 54-59.

\bibitem{hs83} Stapp, H., Exact solution of the infrared problem, {\em Phys. Rev.,} {\bf D28} (1983) 1386-1418.

\bibitem{rfbh18} Flack, R. and  Hiley, B. J.,  Feynman Paths and Weak Values,  in {\em Entropy}, {\bf 20} (5) May 2018.

\bibitem{pd47} Dirac, P. A. M., {\em The Principles of Quantum Mechanics}, Oxford University Press, Oxford, 1947.

\bibitem{wh58} Heisenberg, W., {\em Physics and Philosophy: the revolution in modern science}, George Allen and Unwin, London, 1958.

\bibitem{fbmf78}  Bayen, F., Flato, M., Fronsdal, C., Lichnerowicz, A. and Sternheimer, D., I. Deformation Theory and Quantization of Symplectic Structures, {\em Ann. Phys.} {\bf 111} (1978) 61-110.


\bibitem{tt54} Takabayasi, T., The Formulation of Quantum Mechanics in terms of Ensemble in Phase Space. {\em Prog. Theor. Phy.}, {\bf 11} (4) (1954) 341-373.

\bibitem{rpf48} Feynman, R. P., Space-time Approach to Non-Relativistic Quantum Mechanics, {\em Rev. Mod. Phys}. {\bf 20}, (1948), 367-387.

\bibitem{js51} Schwinger, J., The Theory of Quantum Fields I, {\em Phys. Rev.}, {\bf 82} (1951) 914-927.

\bibitem{js53a} Schwinger, J., The Theory of Quantum Fields II, {\em Phys. Rev.}, {\bf 91} (1953) 713-728.

\bibitem{js53} Schwinger, J., The Theory of Quantum Fields III, {\em Phys. Rev.}, {\bf 91} (1953) 728-740.


\bibitem{pd45} Dirac, P. A. M., On the analogy between Classical and Quantum Mechanics, {\em Rev. Mod. Phys.}, {\bf 17} (1945) 195-199.

\bibitem{db51} Bohm, D., {\em Quantum Theory}, Prentice-Hall, Englewood Cliffs, N.J. (1951).

\bibitem{jm49} Moyal, J. E., Quantum Mechanics as a Statistical Theory, {\em Proc. Camb. Phil. Soc}. {\bf 45} (1949) 99-123.

\bibitem{yalv88} Aharonov, Y., Albert, D. Z. and Vaidman, L., How the Result of a Measurement of a Component of the Spin of a Spin-1/2 Particle Can Turn Out to be 100, {\em Phys. Rev. Lett.}, {\bf 60} (1988) 1351-4.

 \bibitem{rl05} Leavens, C. R., Weak Measurements from the point of view of Bohmian Mechanics, {\em Found. Phys.}, {\bf 35} (2005) 469-91.
 %
\bibitem{hw07} Wiseman, H. M., Grounding Bohmian mechanics in weak values and Bayesianism, {\em New J. Phys}., {\bf 9} (2007) 165-77.

\bibitem{skbbsr11} Kocsis, S., Braverman, B., Ravets, S., Stevens, M. J., Mirin, R. P., Shalm, L.K., Steinberg, A. M., Observing the Average Trajectories of Single Photons in a Two-Slit Interferometer, {\em Science}. {\bf 332} (2011) 1170-73.

\bibitem{cpcd79} Philippidis, C., Dewdney, C. and Hiley, B. J.,   Quantum Interference and the Quantum Potential,   {\em Nuovo Cimento}, {\bf 52B}, 15-28 (1979).

\bibitem{tcrw11} Coffey, T. M. and Wyatt, R. E., Comment on ``Observing the Average Trajectories of Single Photons in a Two-Slit Interferometer", 2011 arXiv:1109.4436.


\bibitem{rfbh16} Flack, R. and  Hiley, B. J.,  Weak Values of Momentum of the Electromagnetic Field: Average Momentum Flow Lines, Not Photon Trajectories. arXiv:1611.06510.

\bibitem{jmpe16} Morley, J., Edmunds, P. D. and Barker, P. F., Measuring the weak value of the momentum in a double slit interferometer, {\em J. Phys. Conference series},   {\bf 701} (2016)  012030.

\bibitem{vmrf17} Monachello, V.,  Flack, R. and  Hiley, B. J.,  A method for measuring the real part of the weak value of spin using non-zero mass particles,  arXiv:1701.04808.

\bibitem{bhj} Born, M., Heisenberg, W. and Jordan, P., On quantum mechanics II, {\em Z. Phys}. {\bf 35} (1926)  557-615.


\bibitem{bh80} Hiley, B. J., Towards an Algebraic Description of Reality, {\em Ann. Fon. Louis de Broglie}, {\bf 5} (1980) 75-103.

\bibitem{bhgd18} Hiley, B. J. and Dennis, G., The Dirac-Bohm Picture, Pre-print available.

\bibitem {bh15} Hiley, B. J.,  On the Relationship between the Moyal Algebra and the Quantum Operator Algebra of von Neumann, {\em Journal of Computational Electronics}, {\bf 14} (2015)  869-878.

\bibitem{od92} Darrigol, O., {\em From c-Numbers to q-Numbers: The Classical Analogy in the History of Quantum Theory}, University of California Press, Berkeley, 1992.

\bibitem{bhbc12} Hiley, B. J. and Callaghan, R. E., Clifford Algebras and the Dirac-Bohm Quantum Hamilton-Jacobi Equation, {\em Found. Phys.}, {\bf 42} (2012) 192-208.

\bibitem {bhdp87} Hiley, B. J. and Peat, D.,  Quantum Implications: Essays in Honour of David Bohm, Routledge \& Kegan Paul, 1987. 












\end{thebibliography}
\bibliographystyle{plain}

\end{document}